\documentstyle[12pt]{article}
\def\beq{\begin{equation}}
\def\eeq{\end{equation}}
\def\be{\begin{equation}}
\def\ee{\end{equation}}
\def\bea{\begin{eqnarray}}
\def\eea{\end{eqnarray}}
\def\p{\partial}
\def\n{\nabla}
\def\d{\displaystyle}

\begin{document}

\thispagestyle{empty}

\begin{titlepage}

\hfill{hep-th/0006144}

\vspace{1cm}

{\baselineskip 20pt

\begin{center}
{\LARGE\bf Causality of Massive Spin 2 Field in External Gravity}

\vspace{1cm}

{\sc I.L. Buchbinder${}^{a,b}$, D.M. Gitman${}^a$
and V.D. Pershin${}^c$
\footnote{e-mail: \tt joseph@tspu.edu.ru, gitman@fma.if.usp.br,
pershin@ic.tsu.ru}

\medskip

\it ${}^a$Instituto de F\'\i sica, Universidade de S\~ao Paulo,\\
\it P.O. Box 66318, 05315-970, S\~ao Paulo, SP, Brasil\\
\it ${}^b$Department of Theoretical Physics,
\it Tomsk State Pedagogical University,\\
\it Tomsk 634041, Russia\\
\it ${}^c$Department of Theoretical Physics, Tomsk State
University,\\ \it Tomsk 634050, Russia }

\end{center}
}

\bigskip

\begin{abstract}
We investigate the structure of equations of motion and lagrangian
constraints in a general theory of massive spin 2 field interacting
with external gravity. We demonstrate how consistency with the flat
limit can be achieved in a number of specific spacetimes. One
such example is an arbitrary static spacetime though equations of
motion in this case may lack causal properties. Another example is
provided by external gravity fulfilling vacuum Einstein equations
with arbitrary cosmological constant. In the latter case there exists
one-parameter family of theories describing causal propagation of the
correct number of degrees of freedom for the massive spin 2 field in
arbitrary dimension. For a specific value of the parameter a gauge
invariance with a vector parameter appears, this value is interpreted
as massless limit of the theory.  Another specific value of the
parameter produces gauge invariance with a scalar parameter and this
cannot be interpreted as a consistent massive or massless theory.
\end{abstract}

\end{titlepage}

Problems of consistent equations of motion for interacting higher
spin fields deserve studying due to many reasons. First of all,
string theory includes an infinite tower of massive excitations with
all possible spins and thus should allow some consistent effective
description of arbitrary spin fields interaction. Second, composite
resonance particles with higher spins do exist and one should be able
to describe their interaction (for example, with external
electromagnetic and gravitational fields) in terms of some effective
local field theory.  At last, investigation of higher spin fields is
interesting on its own from the general point of view. It would be
surprising if nature admits description of free fields with arbitrary
spins but stops, say, at spin 1 in case of interacting massive
fields. Even if it is really the case, one should try to understand
why this is the way the nature works.

This note is devoted to investigation of the massive spin 2 field
interacting with external gravity which represents one of the
simplest higher spin models.  It has been studied in numerous papers
\cite{aragone}-\cite{klish2}%
\footnote{For analysis of the corresponding massless models see e.g.
\cite{vasilev,vasilev2,wald}} but careful and general analysis of the
consistency and causality of the theory in arbitrary curved spacetime
was still absent.

There are at least two ways the interaction may spoil the
consistency of a higher spin theory.
Firstly, interaction may change the number of dynamical
degrees of freedom. For example, a massive field with spin $s$ in
$D=4$ Minkowski spacetime is described by a rank $s$ symmetric
traceless transverse tensor $\phi_{(\mu_1\ldots\mu_s)}$ satisfying
the mass shell condition:
\be
(\p^2-m^2) \phi_{\mu_1\ldots\mu_s} =0 {,} \qquad
\p^\mu \phi_{\mu\mu_1\ldots\mu_{s-1}} =0 {,} \qquad
\phi^\mu{}_{\mu\mu_1\ldots\mu_{s-2}} =0 {.}
\label{irrep}
\ee
To reproduce all these equations from a single lagrangian one needs
to introduce auxiliary fields $\chi_{\mu_1\ldots\mu_{s-2}}$,
$\chi_{\mu_1\ldots\mu_{s-3}}$, \ldots, $\chi$ \cite{fierz,singh}.
These symmetric traceless fields vanish on shell but their presence
in the theory provides lagrangian description of the conditions
(\ref{irrep}).  In higher dimensional spacetimes there appear fields
of more complex tensor structure but general situation remains the
same, i.e.  lagrangian description always requires presence of
unphysical auxiliary degrees of freedom.

Namely these auxiliary fields create problems when one tries to turn
on interaction in the theory. Arbitrary interaction makes the
auxiliary fields dynamical thus increasing the number of degrees of
freedom. Usually these extra degrees of freedom are ghostlike and
should be considered as pathological. Requirement of absence of these
extra dynamical degrees of freedom imposes severe restrictions on the
possible interaction.

To construct a consistent massive field theory one often starts
with a corresponding consistent massless theory which should be
invariant with respect to gauge symmetry and then breaks this
symmetry by introducing mass terms into lagrangian. In this
case invariance of the kinetic part of lagrangian guarantees
the correct number of degrees of freedom in both massless and massive
theories.  For example, spin 2 field possesses a gauge invariant
massless lagrangian only if external gravity satisfies vacuum
Einstein equations and so it is usually believed that the massive
spin 2 field can consistently propagate also only in Einstein
spacetimes.

In this paper which is a sequel to \cite{our2} we show that this
belief is not true and that in massive case there exists a number of
possibilities of providing the correct number of degrees of freedom.
Of course, invariance of kinetic term does always provide the correct
number of degrees of freedom but this is not the only possibility.
In particular, we describe below an example when external spacetime
does not fulfill Einstein equations but the theory is consistent with
the flat spacetime limit.

The other problem that may arise in higher spin fields theories is
connected with possible violation of causal properties. This problem
was first noted in the theory of spin 3/2 field in external fields
\cite{zwanziger} (see also the review \cite{zwanziger2} and a recent
discussion in \cite{3/2})

In general,
when one has a system of differential equations for a set of fields
$\phi^B$ (to be specific, let us say about second order equations)
\be
M_{AB}{}^{\mu\nu}\p_\mu \p_\nu \phi^B + \ldots = 0 {,} \qquad
\mu,\nu=0,\ldots,D-1
\ee
the following definitions are used. A characteristic matrix is the
matrix function of $D$ arguments $n_\mu$ built out of the
coefficients at the second derivatives in the equations:
$
M_{AB}(n) = M_{AB}{}^{\mu\nu} n_\mu n_\nu {.}
$
A characteristic equation is
$
\det M_{AB} (n) = 0 {.}
$
A characteristic surface is the surface $S(x)=const$ where
$\p_\mu S(x)=n_\mu$.

If for any $n_i$ ($i=1,\ldots,D-1$) all solutions of the
characteristic equation $n_0(n_i)$ are real then the system of
differential equations is called hyperbolic and describes
propagation of some wave processes. The hyperbolic system is called
causal if there is no timelike vectors among solutions $n_\mu$ of the
characteristic equations. Such a system describes propagation with a
velocity not exceeding the speed of light. If there exist timelike
solutions for $n_\mu$ then the corresponding characteristic surfaces
are spacelike and violate causality.

Turning on interaction in theories of higher spin fields in general
changes the characteristic matrix and there appears possibility of
superluminal propagation. Such a situation also should be considered
as pathological.

Both these problems arise in the theory of massive
spin 2 field coupled to external gravitational field. To provide
consistency of the interaction we should conserve the same number of
physical degrees of freedom and constraints that the theory possesses
in flat spacetime. To find the complete set of constraints we will
use the general lagrangian scheme \cite{gitman} which
is equivalent to the Dirac-Bergmann procedure in hamiltonian
formalism but for our purposes is simpler.
In the case of second class constraints (which is
relevant for massive higher spin fields) it consists in the following
steps.  If in a theory of some set of fields $\phi^A(x)$,
$A=1,\ldots,N$ the original lagrangian equations of motion define
only $r<N$ of the second time derivatives (``accelerations'')
$\ddot\phi^A$ then one can build $N-r$ primary constraints, i.e.
linear combinations of the equations of motion that does not contain
accelerations. Requirement of conservation in time of the primary
constraints either define some of the missing accelerations or lead
to new (secondary) constraints.  Then one demands conservation of the
secondary constraints and so on, until all the accelerations are
defined and the procedure closes up.

Before considering the theory in external gravitational field we
analyze the structure of equations of motion in Minkowski spacetime.
The purpose of this analysis is twofold. First, we illustrate the
general scheme of calculating the constraints within covariant
lagrangian framework. In addition, building a consistent and causal
theory in curved spacetime we use these flat constraints as a
reference point.

Free spin 2 field is
known to be described by the Fierz-Pauli action \cite{fierz} (we
consider arbitrary spacetime dimension):
\begin{eqnarray}
S&=&\int\! d^D x \biggl\{ \frac{1}{4} \partial_\mu H \partial^\mu H
-\frac{1}{4} \partial_\mu H_{\nu\rho} \partial^\mu H^{\nu\rho}
-\frac{1}{2} \partial^\mu H_{\mu\nu} \partial^\nu H
+\frac{1}{2} \partial_\mu H_{\nu\rho} \partial^\rho H^{\nu\mu}
\nonumber\\&&
\qquad\qquad
{} - \frac{m^2}{4} H_{\mu\nu} H^{\mu\nu} + \frac{m^2}{4}  H^2
 \biggr\} {.}
\label{actfield}
\end{eqnarray}
Here the role of auxiliary field is played by the trace
$H=\eta^{\mu\nu} H_{\mu\nu}$. The equations of motion
\begin{eqnarray}
E_{\mu\nu}&=&\partial^2 H_{\mu\nu} - \eta_{\mu\nu} \partial^2 H +
\partial_\mu \partial_\nu H
+ \eta_{\mu\nu} \partial^\alpha \partial^\beta H_{\alpha\beta}
- \partial_\sigma \partial_\mu H^\sigma{}_\nu
- \partial_\sigma \partial_\nu H^\sigma{}_\mu
\nonumber
\\&&\qquad\qquad
{}-m^2 H_{\mu\nu} + m^2 H \eta_{\mu\nu} = 0
\label{flateq}
\end{eqnarray}
contain $D$ primary constraints (expressions without second time
derivatives $\ddot H_{\mu\nu}$):
\bea
E_{00} &=& \Delta H_{ii} - \p_i\p_j H_{ij} - m^2 H_{ii} \equiv
\varphi_0^{(1)} \approx 0
\\
E_{0i} &=& \Delta H_{0i} +\p_i\dot H_{kk} -\p_k\dot H_{ki} - \p_i\p_k
H_{0k} - m^2 H_{0i} \equiv \varphi_i^{(1)} \approx 0 {.}
\eea
The remaining equations of motion $E_{ij}=0$ allow to define the
accelerations $\ddot H_{ij}$ in terms of $\dot H_{\mu\nu}$ and
$H_{\mu\nu}$. The accelerations $\ddot H_{00}$, $\ddot H_{0i}$ cannot
be expressed from the equations directly.

Conditions of conservation of the primary constraints in time
$\dot E_{0\mu}\approx 0$ lead to $D$ secondary constraints. On-shell
they are equivalent to
\beq
\varphi_\nu^{(2)} = \partial^\mu E_{\mu\nu} =
m^2 \partial_\nu H - m^2 \partial^\mu H_{\mu\nu} \approx 0
\label{con1}
\eeq

Conservation of $\varphi_i^{(2)}$ defines $D-1$ accelerations
$\ddot H_{0i}$ and conservation of $\varphi_0^{(2)}$ gives another
one constraint. It is convenient to choose it in the covariant form
by adding suitable terms proportional to the equations of motion:
\begin{eqnarray}
\varphi^{(3)}=
\partial^\mu \partial^\nu E_{\mu\nu}
+ \frac{m^2}{D-2} \eta^{\mu\nu} E_{\mu\nu}
=  H  m^4 \frac{D-1}{D-2} \approx 0
\label{con2}
\end{eqnarray}
Conservation of $\varphi^{(3)}$ gives one more constraint on initial
values
\beq
\varphi^{(4)} = - \dot H_{00} + \dot H_{kk} = \dot H \approx 0
\eeq
and from the conservation of this last constraint the acceleration
$\ddot H_{00}$ is defined.

Altogether there are $2D+2$ constraints on the initial values of
$\dot H_{\mu\nu}$ and $H_{\mu\nu}$. The lagrangian theory is
equivalent to the system of the equations
\begin{equation}
\Bigr(\partial^2-m^2\Bigl) H_{\mu\nu}=0 {,}\qquad
\partial^\mu H_{\mu\nu}=0 {,}\qquad
H^\mu{}_\mu=0 {.}
\label{irred}
\end{equation}
and describes traceless and transverse symmetric tensor field of the
second rank.

Obviously, the equations of motion (\ref{irred}) are causal because
the characteristic equation
\be
\det M(n) = (n^2)^{D(D+1)/2}
\ee
has 2 multiply degenerate roots
\be
-n_0^2+n_i^2 = 0, \qquad n_0 = \pm \sqrt{n_i^2} {.}
\ee
which correspond to real null solutions for $n_\mu$. Note that
analysis of causality is possible only after calculation of all the
constraints. Original lagrangian equations of motion (\ref{flateq})
have degenerate characteristic matrix $\det M(n)\equiv 0$ and do not
allow to define propagation cones of the field $H_{\mu\nu}$.

In the massless limit $m^2=0$ the structure of the theory
(\ref{actfield}) changes. Instead of the secondary constraints
(\ref{con1}) conservation of the primary constraints lead to
identities $\p^\mu E_{\mu\nu} \equiv 0$ which mean that the theory
becomes gauge invariant with respect to the local transformations
$\delta H_{\mu\nu} = \p_\mu \xi_\nu + \p_\nu \xi_\mu$. Such a theory
represents the quadratic part of the Einstein-Hilbert action for
gravitational field and the gauge invariance is a linear counterpart
of the general coordinate invariance.

Now if we want to construct a theory of massive spin 2 field on a
curved manifold first of all we should provide the same number of
propagating degrees of freedom as in the flat case. It means that new
equations of motion $E_{\mu\nu}$ should lead to exactly $2D+2$
constraints  and in the flat spacetime limit these constraints should
reduce to their flat counterparts.

Generalizing (\ref{actfield}) to curved spacetime we should substitute
all derivatives by the covariant ones and also we can add
non-minimal terms containing curvature tensor with some
dimensionless coefficients in front of them.  As a result, the most
general action for massive spin 2 field in curved spacetime quadratic
in derivatives and consistent with the flat limit should have the
form \cite{aragone}:
\begin{eqnarray}&&
S=\int d^D x\sqrt{-G} \biggl\{ \frac{1}{4} \nabla_\mu H \nabla^\mu H
-\frac{1}{4} \nabla_\mu H_{\nu\rho} \nabla^\mu H^{\nu\rho}
-\frac{1}{2} \nabla^\mu H_{\mu\nu} \nabla^\nu H
+\frac{1}{2} \nabla_\mu H_{\nu\rho} \nabla^\rho H^{\nu\mu}
\nonumber
\\&&
\qquad
{}+\frac{a_1}{2} R H_{\alpha\beta} H^{\alpha\beta}
+\frac{a_2}{2} R H^2
+\frac{a_3}{2} R^{\mu\alpha\nu\beta} H_{\mu\nu} H_{\alpha\beta}
+\frac{a_4}{2} R^{\alpha\beta} H_{\alpha\sigma} H_\beta{}^\sigma
+\frac{a_5}{2} R^{\alpha\beta} H_{\alpha\beta} H
\nonumber
\\&&
\qquad
{}- \frac{m^2}{4} H_{\mu\nu} H^{\mu\nu} + \frac{m^2}{4} H^2
 \biggr\}
\label{genact}
\end{eqnarray}
where $a_1, \ldots a_5$ are so far arbitrary dimensionless
coefficients, $R^\mu{}_{\nu\lambda\kappa}=\partial_\lambda
\Gamma^\mu_{\nu\kappa} -\ldots$,
$R_{\mu\nu}=R^\lambda{}_{\mu\lambda\nu}$.

Equations of motion
\begin{eqnarray}
E_{\mu\nu}&=&\nabla^2 H_{\mu\nu} - G_{\mu\nu} \nabla^2 H +
\nabla_\mu \nabla_\nu H
+ G_{\mu\nu} \nabla^\alpha \nabla^\beta H_{\alpha\beta}
- \nabla_\sigma \nabla_\mu H^\sigma{}_\nu
- \nabla_\sigma \nabla_\nu H^\sigma{}_\mu
\nonumber
\\&&
{}+2a_1 R H_{\mu\nu}
+2a_2 G_{\mu\nu} R H
+2a_3 R_\mu{}^\alpha{}_\nu{}^\beta H_{\alpha\beta}
+a_4 R_\mu{}^\alpha H_{\alpha\nu}
+a_4 R_\nu{}^\alpha H_{\alpha\mu}
\nonumber
\\&&
{}+a_5 R_{\mu\nu} H
+a_5 G_{\mu\nu} R^{\alpha\beta} H_{\alpha\beta}
-m^2 H_{\mu\nu} + m^2 H G_{\mu\nu} \approx 0
\label{lagreq}
\end{eqnarray}
contain second time derivatives of $H_{\mu\nu}$ in the following
way:
\bea
E_{00} &=& (G^{mn}-G_{00}G^{00}G^{mn}+G_{00}G^{0m}G^{0n}) \nabla_0
\nabla_0 H_{mn} + O(\nabla_0) {,}
\nonumber\\
E_{0i} &=&
(-G_{0i}G^{00}G^{mn} + G_{0i}G^{0m}G^{0n} - G^{0m}\delta^n_i)
\nabla_0 \nabla_0 H_{mn} + O(\nabla_0) {,}
\nonumber\\
E_{ij} &=&
(G^{00}\delta^m_i\delta^n_j - G_{ij}G^{00}G^{mn} +
G_{ij}G^{0m}G^{0n}) \nabla_0 \nabla_0 H_{mn} + O(\nabla_0) {.}
\eea
So we see that accelerations $\ddot H_{00}$ and $\ddot H_{0i}$ again
(as in the flat case) do not enter the equations of motion while
accelerations $\ddot H_{ij}$ can be expressed through $\dot
H_{\mu\nu}$, $H_{\mu\nu}$ and their spatial derivatives.

There are $D$ linear combinations of the equations of motion which
do not contain second time derivatives and so represent primary
constraints of the theory:
\be
\varphi^{(1)}_\mu = E^0{}_\mu = G^{00} E_{0\mu} + G^{0j} E_{j\mu}
\label{phi1}
\ee
Now one should calculate time derivatives of these constraints and
define secondary ones. In order to do this in a covariant form
we can add to the time derivative of $\varphi^{(1)}_\mu$ any linear
combination of equations of motion and primary constraints. So we
choose the secondary constraints in the following way:
\bea
\varphi^{(2)}_\mu &=& \nabla^\alpha E_{\alpha\mu} =
\dot\varphi^{(1)}_\mu + \p_i E^i{}_\mu
+ \Gamma^\alpha_{\alpha0}\varphi^{(1)}_\mu
+ \Gamma^\alpha_{\alpha i} E^i{}_\mu
- \Gamma^\sigma_{\mu 0}\varphi^{(1)}_\sigma
- \Gamma^\sigma_{\mu i} E^i{}_\sigma
\nonumber\\
&=&
(2a_1 R-m^2) \nabla^\mu H_{\mu\nu}
+(2a_2 R+m^2) \nabla_\nu H
+2a_3 R^{\mu\alpha}{}_\nu{}^\beta \nabla_\mu H_{\alpha\beta}
+a_4 R^{\mu\alpha} \nabla_\mu H_{\alpha\nu}
\nonumber\\&&{}
+(a_4-2) R^\alpha{}_\nu \nabla^\mu H_{\alpha\mu}
+a_5 R^{\alpha\mu} \nabla_\nu H_{\alpha\mu}
+(a_5+1) R^\alpha{}_\nu \nabla_\alpha H
\nonumber\\&&{}
+ (2a_1+\frac{a_4}{2}) H_{\alpha\nu} \nabla^\alpha R
+ (2a_2+\frac{a_5}{2}) H \nabla_\nu R
\nonumber\\&&{}
+ H_{\alpha\beta} \biggr[ (2a_3+a_5+1) \nabla_\nu R^{\alpha\beta}
+ (a_4-2a_3-2) \nabla^\alpha R^\beta{}_\nu \biggl]
\label{phi2}
\eea
At the next step conservation of these $D$ secondary constraints
should lead to one new constraint and to expressions for $D-1$
accelerations $\ddot H_{0i}$. This means that the constraints
(\ref{phi2}) should contain the first time derivatives $\dot
H_{0\mu}$ through the matrix with the rank $D-1$:
\bea
\varphi^{(2)}_0 &=&
A \; \dot H_{00} + B^j \dot H_{0j} + \ldots
\nonumber\\
\varphi^{(2)}_i &=&
C_i \dot H_{00} + D_i{}^j \dot H_{0j} + \ldots
\label{velocities}
\eea
\be
\mbox{rank}\; \hat\Phi_\mu{}^\nu \equiv
\mbox{rank} \left|\left|
\begin{array}{cc}
A& B^j \\ C_i & D_i{}^j
\end{array}
\right|\right| = D-1
\label{rank}
\ee
In the flat spacetime we had the matrix
\be
\hat\Phi_\mu{}^\nu =
\left|\left|
\begin{array}{cc}
0& 0 \\ 0 & m^2 \delta_i^j
\end{array}
\right|\right|
\ee
In the curved case the explicit form of this matrix elements in the
constraints (\ref{phi2}) is:
\bea
A &=& R G^{00} (2a_1+2a_2) + R^{00} (a_4+a_5)
           + R^0{}_0 G^{00} (a_4+a_5-1)
\nonumber\\
B^j &=& m^2 G^{0j} + R G^{0j} (2a_1+4a_2) + 2a_3 R^{0j}{}_0{}^0
      + R^j{}_0 G^{00} (a_4-2)
\nonumber\\&&{}
      + R^{0j} (a_4+2a_5) + R^0{}_0 G^{0j} (a_4+2a_5)
\nonumber\\
C_i &=& R^0{}_i G^{00} (a_4+a_5-1)
\nonumber\\
D_i{}^j &=&{} - m^2 G^{00} \delta_i^j + 2a_1 RG^{00} \delta_i^j
 + 2a_3 R^{0j}{}_i{}^0 + a_4 R^{00} \delta_i^j
\nonumber\\&&{}
 + (a_4-2) R^j{}_i G^{00} + (a_4+2a_5) R^0_i G^{0j}
\label{elements}
\eea

At this stage the restrictions that consistency imposes on the
type of interaction reduce to the requirements that the above matrix
elements give
\be
\det\hat\Phi=0 {,} \qquad \det D_i{}^j \neq 0
\label{main}
\ee
When the gravitational background is arbitrary it is not
clear how to fulfill this condition by choosing some
specific values of non-minimal couplings $a_1$, \ldots $a_5$. For
example, requirement of vanishing of the elements $A$
and $C_i$ (\ref{elements}) would lead to contradictory equations
$a_4+a_5=0$, $a_4+a_5-1=0$.

But the consistency conditions (\ref{main}) can be fulfilled in a
number of specific gravitational background. Namely, any spacetime
which in some coordinates has
\be
R^0{}_i=0
\label{example}
\ee
provides such an example.
In such a spacetime $R^{00}=R^0{}_0G^{00}$ and choosing coefficients
$a_1+a_2=0$, $2a_4+2a_5=1$ we have the first column of the matrix
$\hat\Phi$ vanishing and so the conditions (\ref{main}) fulfilled.

As a first example where (\ref{example}) holds let us consider an
arbitrary static spacetime, i.e. a spacetime having a timelike
Killing vector and invariant with respect to the time reversal
$x^0\to -x^0$. In such a spacetime one can always find
coordinates where
\be
\p_0 G_{\mu\nu}=0 {,} \qquad  G_{0i}=0 {.}
\ee
The matrix elements (\ref{elements}) in this case become
\bea
&& A=RG^{00} (2a_1+2a_2) + R^{00} (2a_4+2a_5-1) {,}
\qquad B^j = 0 {,}
\qquad C_i = 0 {,}
\nonumber\\
&& D_i{}^j = ({} -m^2 G^{00} + 2a_1 RG^{00} + a_4 R^{00}) \delta_i^j
 + (a_4-2) R^j{}_i G^{00} + 2a_3 R^{0j}{}_i{}^0
\eea
and (\ref{main}) lead to the following conditions:
\be
2a_1+2a_2=0,  \qquad 2a_4+2a_5-1 =0 , \qquad
\det D_i{}^j \neq 0
\label{eqineq}
\ee
The last inequality may be violated in strong gravitational field and
as we comment below this fact may lead to causal problems.

Suppose that all the conditions (\ref{eqineq}) are fulfilled.
For simplicity we also choose $a_3=0$. Then we have the equations of
the form (\ref{lagreq}) with the coefficients
\be
a_1=\frac{\xi_1}{2}, \quad
a_2=-\frac{\xi_1}{2}, \quad
a_3=0, \quad
a_4=\frac{1}{2}-\xi_2, \quad
a_5=\xi_2
\ee
where $\xi_1$, $\xi_2$ are two arbitrary coupling parameters.

One of the secondary constraints
\bea
\varphi^{(2)}_0 &=& \n^\alpha E_{\alpha 0} =
\n_0 H_{ij} \biggl[ G^{ij} (m^2-\xi_1 R) + (1+\xi_2) G^{ij} R^0{}_0
+\xi_2 R^{ij} \biggr]
\nonumber\\
&&{}
+ \n_i H_{0j} \biggl[ G^{ij} (\xi_1 R-m^2)
- \biggl(\frac{3}{2}+\xi_2\biggr) G^{ij} R^0{}_0
+ \biggl(\frac{1}{2}-\xi_2\biggr) R^{ij} \biggr]
\nonumber\\
&&{}
+ \biggl( \frac{1}{4}+\xi_1-\frac{\xi_2}{2} \biggr) H_{0i}\n^i R
- \biggl( \frac{3}{2}+\xi_2 \biggr) H_{0i}\n^i R^0{}_0
\eea
does not contain velocities $\dot H_{00}$, $\dot H_{0i}$ and so
its conservation leads to a new constraint
$\varphi^{(3)} \approx \n_0\n^\alpha E_{\alpha 0}$. After exclusion
from this expression the accelerations $\ddot H_{ij}$ we get this
constraint as the following combination of the equations of motion:
\bea
\varphi^{(3)} &=& \n_0\n^\mu E_{\mu0} - \xi_2 G_{00} R^{ij} E_{ij}
+\frac{1}{D-2} \Bigl[ m^2 G_{00} + (\xi_2-\xi_1)RG_{00} + R_{00}\Bigr]
G^{ij} E_{ij}
\nonumber\\
&=& \biggl\{ \frac{D-1}{D-2} m^4 +
\frac{2\xi_2-2\xi_1(D-1)}{D-2} m^2 R + \biggl(
2\xi_2+\frac{D-1}{D-2}\biggr) m^2 R^0{}_0
-\xi_2^2 R_{ij} R^{ij}
\nonumber\\&&{}
 \!\! + \xi_1^2 RR +
\biggl( \frac{\xi_2-\xi_1(D-1)}{D-2} -2\xi_1\xi_2\biggr) RR^0{}_0
+ \xi_2(\xi_2+1) R_{00} R^{00} \biggr\} H_{00} + \ldots
\eea

We did not write down the explicit form of this constraint because
everything we should know about it is the way it contains the
component $H_{00}$. Namely, $\varphi^{(3)}$  contains neither the
acceleration $\ddot H_{00}$ nor the velocity $\dot H_{00}$. It means
that its conservation in time leads to another new constraints
\be
\varphi^{(4)}\approx \n_0 \varphi^{(3)}
\ee
and hence the total number of constraints is the
same as in the flat spacetime provided that the expression in the
braces in front of $H_{00}$ in $\varphi^{(3)}$ does not vanish.

Vanishing of this expression in braces as well as violation of the
inequality (\ref{eqineq}) leads to local changing of the number of
degrees of freedom and this fact is known to be related with acausal
behavior in higher spin theories in external fields \cite{3/2,g=2}.
In general, causality breaks in those cases when there are points
in spacetime in which it is impossible to define all the
accelerations from the conservation of constraints.

In our case it means that causality will hold everywhere only if
$\det D^i{}_j\neq 0$ and the expression in braces in $\varphi^{(3)}$
also does not vanish. Obviously, in general case there are values of
$R_{\mu\nu}$ that violate these requirements.

It is instructive to consider in more detail the Reissner-Nordstrom
solution in $D=4$ as a simple example of non-trivial static spacetime
\be
ds^2 = - (1-\frac{2M}{r}+\frac{Q^2}{r^2}) dt^2
+ \frac{dr^2}{1-\frac{2M}{r}+\frac{Q^2}{r^2}} + r^2 d\Omega^2
\ee
In this case causality problems are absent when the expressions
\bea
\det D_i{}^j &\sim& \biggl( m^2-(1+2\xi_2)\frac{Q^2}{r^4}\biggr)
\biggl( m^2+2\frac{Q^2}{r^4}\biggr)^2
\nonumber\\
\varphi^{(3)} &=& \biggl\{ \frac{D-1}{D-2} m^4 +
m^2 \biggl(2\xi_2+\frac{D-1}{D-2}\biggr)\frac{Q^2}{r^4}
+\xi_2 (1-2\xi_2) \frac{Q^4}{r^8}
\biggr\} H_{00} + \ldots
\label{zeroes}
\eea
do not vanish. Far enough from the horizon where all the terms
containing $r$ in (\ref{zeroes}) are too small and so propagation is
causal in this region. Causal problems may develop only for small
values of $r$. This might be excluded if all terms in (\ref{zeroes})
were positive, that is if
\be
1+2\xi_2 < 0, \qquad 2\xi_2+\frac{D-1}{D-2} >0, \qquad
\xi_2 (1-2\xi_2) > 0
\ee
but these three conditions are contradictory.

It means that for any value of the coupling parameter expressions in
(\ref{zeroes}) vanish for some values of the coordinate $r$ and
the massive spin 2 field propagate causally only in the regions near
infinity but close to the horizon causality is lost. Of course,
this example does not mean that there cannot exist other spacetimes
where causality might be achieved everywhere for some special values
of coupling parameters.

Another possible way to fulfill the consistency requirements
(\ref{main}) is to consider spacetimes representing solutions
of vacuum Einstein equations with arbitrary cosmological
constant:
\be
R_{\mu\nu}=\frac{1}{D}G_{\mu\nu}R \; {.}
\label{einst}
\ee
In this case the coefficients $a_4$, $a_5$ in the lagrangian
(\ref{genact}) are absent and the matrix $\hat\Phi$ takes the form:
\be
\hat\Phi_\mu{}^\nu=
\left|\left|
\begin{array}{c|c}
R G^{00}(2a_1+2a_2-\frac{\d 1}{\d D})
& R G^{0j}(2a_1+4a_2) + 2a_3R^{0j}{}_0{}^0 + m^2 G^{0j} \\
~&~\\
\hline
~&~\\
0& 2a_3 R^{0j}{}_i{}^0 +
 R G^{00} \delta_i^j (2a_1-\frac{\d 2}{\d D}) - m^2 G^{00} \delta_i^j
\end{array}
\right|\right|
\ee
The simplest way to make the rank of this matrix to be equal to $D-1$
is provided by the following choice of the coefficients:
\be
2a_1 + 2a_2 -\frac{1}{D} = 0, \qquad a_3 = 0, \qquad
2R \biggl(a_1-\frac{1}{D}\biggr) - m^2 \neq 0 {.}
\ee

As a result, we have one-parameter family of theories:
\bea
&&
a_1 =\frac{\xi}{D}, \quad a_2 = \frac{1-2\xi}{2D},\quad
a_3=0,\quad a_4=0, \quad a_5 = 0
\nonumber\\&&
R_{\mu\nu}=\frac{1}{D}G_{\mu\nu}R , \qquad
\frac{2(1-\xi)}{D} R + m^2 \neq 0 {.}
\eea
with $\xi$ an arbitrary real number.

The action in this case takes the form
\bea
&&
S=\int d^D x\sqrt{-G} \biggl\{ \frac{1}{4} \nabla_\mu H \nabla^\mu H
-\frac{1}{4} \nabla_\mu H_{\nu\rho} \nabla^\mu H^{\nu\rho}
-\frac{1}{2} \nabla^\mu H_{\mu\nu} \nabla^\nu H
+\frac{1}{2} \nabla_\mu H_{\nu\rho} \nabla^\rho H^{\nu\mu}
\nonumber
\\&&
\qquad\qquad
+\frac{\xi}{2D} R H_{\mu\nu} H^{\mu\nu} +\frac{1-2\xi}{4D} R H^2
- \frac{m^2}{4} H_{\mu\nu} H^{\mu\nu} + \frac{m^2}{4} H^2
 \biggr\} {.}
\label{curvact}
\eea
and the corresponding equations of motion are
\bea
&&
E_{\mu\nu}=\nabla^2 H_{\mu\nu} - G_{\mu\nu} \nabla^2 H +
\nabla_\mu \nabla_\nu H
+ G_{\mu\nu} \nabla^\alpha \nabla^\beta H_{\alpha\beta}
- \nabla_\sigma \nabla_\mu H^\sigma{}_\nu
- \nabla_\sigma \nabla_\nu H^\sigma{}_\mu
\nonumber
\\&&\qquad\qquad
+ \frac{2\xi}{D} R H_{\mu\nu} + \frac{1-2\xi}{D} RH G_{\mu\nu}
-m^2 H_{\mu\nu} + m^2 H G_{\mu\nu} = 0
\label{cons_eq}
\eea
The secondary constraints built out of them are
\be
\varphi^{(2)}_\mu=\nabla^\alpha E_{\alpha\mu} =
(\nabla_\mu H -\nabla^\alpha H_{\mu\alpha})
\biggl( m^2+\frac{2(1-\xi)}{D}R\biggr)
\label{seccon}
\ee
and the matrix $\hat\Phi$ looks like
\be
\hat\Phi_\mu{}^\nu =
\biggl( m^2+\frac{\d 2(1-\xi)}{\d D}R\biggr)
\left|\left|
\begin{array}{c|c}
0& G^{0j}  \\
\hline
0 & - G^{00}\delta_i^j
\end{array}
\right|\right|
\ee
Just like in the flat case, in this theory the conditions
$\dot \varphi^{(2)}_i\approx 0$ define the accelerations $\ddot H_{0i}$
and the condition $\dot\varphi^{(2)}_0\approx 0$ after excluding
$\ddot H_{0i}$ gives a new constraint, i.e. the acceleration $\ddot
H_{00}$ is not defined at this stage.

To define the new constraint in a covariant form we use the
following linear combination of $\dot\varphi^{(2)}_\mu$, equations of
motion, primary and secondary constraints:
\bea
\varphi^{(3)} &=&
\frac{m^2}{D-2} G^{\mu\nu} E_{\mu\nu}
+ \nabla^\mu \nabla^\nu E_{\mu\nu}
+ \frac{2(1-\xi)}{D(D-2)} R G^{\mu\nu} E_{\mu\nu} =
\nonumber
\\&=& H \frac{1}{D-2} \biggl( \frac{2(1-\xi)}{D} R + m^2 \biggr)
\biggl( \frac{D+2\xi(1-D)}{D} R + m^2 (D-1)\biggr) \approx 0 {.}
\eea
This gives tracelessness condition for the field $H_{\mu\nu}$
provided that parameters of the theory fulfill the conditions:
\be
\frac{2(1-\xi)}{D} R + m^2 \neq 0 {,} \qquad
\frac{D+2\xi(1-D)}{D} R + m^2 (D-1) \neq 0
\label{ineq}
\ee

Requirement of conservation of $\varphi^{(3)}$ leads to one more
constraint
\bea
\dot \varphi^{(3)} \sim \dot H  \quad\Longrightarrow\quad
\varphi^{(4)} = \dot H \approx 0 {.}
\eea
The last acceleration $\ddot H_{00}$ is expressed from the
condition $\dot\varphi^{(4)}\approx 0$.

Using the constraints for simplifying the equations of motion we see
that the original equations are equivalent to the following system:
\bea
&&\nabla^2 H_{\mu\nu}
+ 2 R^\alpha{}_\mu{}^\beta{}_\nu H_{\alpha\beta}
+ \frac{2(\xi-1)}{D} R H_{\mu\nu} - m^2 H_{\mu\nu} = 0 {,}
\nonumber\\&&
H^\mu{}_\mu=0 {,} \qquad\qquad\dot H^\mu{}_\mu=0 {,} \qquad\qquad
\nabla^\mu H_{\mu\nu} = 0 {,}
\label{curv_shell}
\\&&
G^{00}\n_0\n_i H^i{}_\nu - G^{0i}\n_0\n_i H^0{}_\nu
- G^{0i}\n_i\n_0 H^0{}_\nu - G^{ij}\n_i\n_j H^0_\nu
-2R^{\alpha0\beta}{}_\nu H_{\alpha\beta}
\nonumber
\\&&{}\qquad\qquad\qquad\qquad
- \frac{2(\xi-1)}{D} RH^0{}_\nu + m^2 H^0{}_\nu = 0 {.}
\nonumber
\eea
The last expression represents $D$ primary constraints.

For any values of $\xi$ (except two degenerate values excluded by
(\ref{ineq})) the theory describes the same number of degrees of
freedom as in the flat case - the symmetric, covariantly transverse
and traceless tensor.  $D$ primary constraints guarantees
conservation of the transversality conditions in time.

Let us now consider the causal properties of the theory. Again, if we
tried to use the equations of motion in the original lagrangian form
(\ref{cons_eq}) then the characteristic matrix
\be
M_{\mu\nu}{}^{\lambda\kappa} (n) =
\delta_{(\mu\nu)}{}^{(\lambda\kappa)} n^2
- G_{\mu\nu} G^{\lambda\kappa}n^2
+ G^{\lambda\kappa} n_\mu n_\nu
+ G_{\mu\nu} n^\lambda n^\kappa
- \delta_\nu^{(\kappa} n^{\lambda)} n_\mu
- \delta_\mu^{(\kappa} n^{\lambda)} n_\nu
\ee
would be degenerate. This fact can be seen from the relation
\be
n^\mu M_{\mu\nu}{}^{\lambda\kappa} (n) \equiv 0
\ee
which means that any symmetric tensor of the form $n_{(\mu}t_{\nu)}$
(with $t_\nu$ an arbitrary vector) represents a ``null vector'' for
the matrix $M(n)$  and therefore $\det M=0$.

After having used the constraints we obtain the equations of motion
written in the form (\ref{curv_shell}) and the characteristic
matrix becomes non-degenerate:
\be
M_{\mu\nu}{}^{\lambda\kappa} (n) =
\delta_{\mu\nu}{}^{\lambda\kappa} n^2, \qquad
n^2 = G^{\alpha\beta} n_\alpha n_\beta {.}
\ee
The characteristic cones remains the same as in the flat case. At any
point $x_0$ we can choose locally
$G^{\alpha\beta}(x_0)=\eta^{\alpha\beta}$
and then
\be
\left. n^2 \right|_{x_0} = - n_0^2 + n_i^2
\ee
Just like in the flat case the equations are hyperbolic and causal.

Now let us discuss the massless limit of the theory under
consideration. There are several points of view on the
definition of masslessness in a curved spacetime of an arbitrary
dimension. We guess that the most physically accepted definition is
the one referring to appearance of a gauge invariance for some
specific values of the theory parameters (see e.g.
\cite{massless,nolland} for a recent discussion).

In our case it means that the real mass parameter $M$
for the field $H_{\mu\nu}$ in an Einstein spacetime is defined as
\be
M^2 = m^2 + \frac{2(1-\xi)}{D} R
\label{realm}
\ee
When $M^2=0$ instead of $D$ secondary constraints $\varphi_\mu^{(2)}$
we have $D$ identities for the equations of motion
$\n^\mu E_{\mu\nu}\equiv 0$ and the theory acquires gauge invariance
$\delta H_{\mu\nu}=\n_\mu \xi_\nu + \n_\nu \xi_\mu$. This explains
the meaning of the first condition in (\ref{ineq}), it just tells us
that the theory is massive.

In fact, two parameters $m^2$ and $\xi$
enter the action (\ref{curvact}) in a single combination $M^2$
(\ref{realm}). Since scalar curvature is constant in Einstein
spacetime there is no way to distinguish between the corresponding
terms $\sim \xi R HH$, $\sim m^2 HH$ (with arbitrary $\xi$, $m$) in
the action.  The difference between the two will appear only if we
consider Weyl rescaling of the metric. Note that the ``massless''
theory with $M^2=0$ is not Weyl invariant. In the case of dS/AdS
spacetimes the difference between masslessness, conformal and gauge
invariance and null cone propagation was discussed in detail in
\cite{desnep}.  In our case the theory obviously cannot possess Weyl
invariance.

The second inequality (\ref{ineq}) is more mysterious. If it fails to
hold, i.e. if
$
M^2 = M^2_c \equiv \frac{D-2}{D(D-1)} R
$
then instead of the constraint $\varphi^{(3)}$ the scalar identity
\be
\n^\mu\n^\nu E_{\mu\nu} + \frac{R}{D(D-1)} G^{\mu\nu} E_{\mu\nu} =0
\ee
with the corresponding gauge invariance
\be
\delta H_{\mu\nu} = \n_\mu \n_\nu \epsilon + \frac{R}{D(D-1)}
G_{\mu\nu} \epsilon
\label{strange}
\ee
arise.

Appearance of this gauge invariance with a scalar parameter was first
found for the massive spin 2 in spacetime of constant curvature in
\cite{desnep} and was further investigated \cite{higuchi,bengt} in
spacetimes with positive cosmological constant. Our analysis shows
that this gauge invariance is a feature of more general spin 2
theories in arbitrary Einstein spacetimes.  In this case we can
simplify the equations of motion using the secondary constraints
(\ref{seccon}):
\be
\n^2 H_{\mu\nu}- \n_\mu\n_\nu H
+ 2R_\mu{}^\alpha{}_\nu{}^\beta H_{\alpha\beta}
+ \frac{2-D}{D(D-1)} RH_{\mu\nu} -\frac{1}{D(D-1)}
RG_{\mu\nu} H = 0 {.}
\ee
After imposing the gauge condition\footnote{It does not fix
(\ref{strange}) completely and the residual symmetry
with the prameter obeying
$\bigl(\n^2+\frac{R}{D-1}\bigr)\epsilon=0$ remains.} $H=0$ one can see
that these equations describe causal propagation of the field
$H_{\mu\nu}$ but the number of propagating degrees of freedom
corresponds to neither massive nor massless spin 2 free field.
It was argued in \cite{higuchi,bengt} that appearance of the gauge
invariance (\ref{strange}) leads to such
pathological properties as  violation of the classical Hamiltonian
positiveness and negative norm states in the quantum version of the
theory. One should expect similar problems in the general
spin 2 theory in arbitrary Einstein spacetime described in this
paper.

\medskip

We demonstrated that correct number of degrees of freedom in the
massive spin 2 theory (algebraic consistency) can be achieved in a
large class of curved spacetimes which include as particular cases
arbitrary static spacetimes and vacuum Einstein spacetimes. An
analysis of the constraints structure shows that in case of a static
spacetime there esxists a potential source of acausal behavior.
However, as we see on the example of Reissner-Norsdtrom spacetime
causal propagation is possible in the regions where gravitational
field is weak enough. In Einstein spacetimes spin 2 massive field can
be consistently described by a one-parameter family of theories
(\ref{curvact}). For any value of the parameter satisfying
(\ref{ineq}) the corresponding equations describe the correct number
of degrees of freedom which propagate causally.

It is interesting to compare our approach with the paper \cite{CD}
which is devoted to investigation of consistency of higher rank
spin-tensor fields in curved spacetime from a different point of view.
The authors of \cite{CD} considered equations for the fields carrying
irreducible representations of Euclidian version of the four
dimensional Lorentz group SO(4) and analyzed when irreducibility of
these representations is preserved in curved space.  In particular,
they showed that symmetric second rank tensor equations are
consistent in this sense in Einstein spaces.  However, such an
analysis (sufficient for the proof of index theorems in \cite{CD}) is
not enough when one tries to build a consistent theory for a
physical field on a curved manifold starting from an irreducible
representation of the Poincare group in flat spacetime. Preservation
of the correct number of degrees of freedom in such a theory is a
requirement independent from the algebraic consistency considered in
\cite{CD}.  Einstein spacetimes provide an example when both these
conditions are fulfilled but as we saw in our analysis correct number
of degrees of freedom can be preserved in much wider class of
spacetimes. Besides, in physical theories for interacting higher spin
fields we face a new problem of causality which should also be
studied independently. In general there can be theories with correct
number of degrees of freedom but acausal, and we really see such
examples in case of spin 2 field in external gravity. It is worth to
note that the authors of \cite{CD} emphasized that their analysis has
no direct relation to the problem of consistent propagation of higher
spin physical fields and that they did not set this problem at all.

In case of Einstein spacetimes our lagrangian for the spin 2 field in
curved spacetime is the most general known so far, in all previous
works only the theories with specific values of the parameter $\xi$
were considered \cite{bengt,ovrut}. Two degenerate values of the
parameter $\xi$ describe the theories with different degrees of
freedom. One of this degenerate values corresponds to massless spin 2
field in an Einstein spacetime, another one describes neither massive
nor massless spin 2 field.

The next natural step would consist in building a theory describing
dynamics of both gravity and massive spin 2 field. In such a theory
in addition to dynamical equations for the massive spin 2 field one
would have dynamical equations for gravity with the energy-momentum
tensor constructed out of spin 2 field components.  The analysis of
consistency then changes and one needs to have correct number of
constraints and causality for both fields interacting with each
other \cite{aragone}.

The only known consistent system of a higher spin field interacting
with dynamical gravity is the theory of massless helicity 3/2 field,
i.e. supergravity \cite{sugra} (see also the book \cite{book}). In
that case consistency with dynamical gravity requires four-fermion
interaction. If a consistent description of spin 2 field interacting
with dynamical gravity exists it may also require some non-trivial
modification of the lagrangian.  At least, it is known that
lagrangians quadratic in spin 2 field do not provide such a
consistency \cite{aragone}. A possible way of consistent
description of the spin 2 field on arbitrary gravitational background
was recently proposed in \cite{our2}. This was achieved by
means of representation of the lagrangian in the form of infinite
series in curvature and imposing the consistency condition
perturbatively in each order (earlier similar construction was
investigated for symmetric Einstein spacetime in \cite{klish2}).

Further generalizations of our analysis may include theories of
massive spins $s\ge3$ fields (which would require more complex
structure of auxiliary fields) and interaction with other background
fields, e.g.  with scalar dilaton and antisymmetric tensor
that are relevant in string theory.

\medskip

{\bf Acknowledgements.}
We are grateful to V.~Krykhtin, S.~Kuzenko, H.~Osborn, B.~Ovrut,
A.~Tseytlin, M.~Vasiliev and G.~Veneziano for useful discussions of
some aspects of this work.  The work of I.L.B. and V.D.P. was
supported by GRACENAS grant, project 97-6.2-34 and RFBR grant,
project 99-02-16617; the work of I.L.B. and V.D.P. was supported by
RFBR-DFG grant, project 99-02-04022 and INTAS grant N 991-590. I.L.B.
is grateful to FAPESP and D.M.G. is grateful to CNPq for support of
the research.

\end{document}